\documentclass{elsart}

\usepackage{graphicx}

\usepackage{amssymb}

\begin{document}

\begin{frontmatter}



\title{Faddeev-Merkuriev equations for resonances 
in three-body Coulombic systems \thanksref{koszi}}

\thanks[koszi]{This work is dedicated to the
70th birthday of Prof.\ Borb\'ala Gyarmati}


\author[label1,label2]{Z.~Papp}
\author[label3]{J.~Darai}
\author[label2]{A.~Nishimura}
\author[label2]{Z.~T.~Hlousek}
\author[label2]{C.-Y.~Hu}
\author[label4]{S.~L.~Yakovlev}

\address[label1]{
Institute of Nuclear Research of the Hungarian 
         Academy of Sciences,
         PO Box 51, H--4001 Debrecen, Hungary}

\address[label2]{ Department of Physics, 
California State University, Long Beach, California 90840}

\address[label3]{ Department of Experimental Physics, University of Debrecen, 
 Debrecen, Hungary 
}

\address[label4]{ Department of Mathematical and Computational Physics, 
St.\ Petersburg State University, St.\ Petersburg,  Russia }

\begin{abstract}
We reconsider the homogeneous Faddeev-Merkuriev integral equations for
three-body Coulombic systems with attractive Coulomb interactions 
and point out that the resonant solutions are
contaminated with spurious resonances. The spurious solutions are
related to the splitting of the attractive Coulomb potential into 
short- and long-range parts, which is inherent in the approach, but arbitrary
to some extent. By varying the parameters of the splitting the spurious
solutions can easily be ruled out. We solve the integral equations by 
using the Coulomb-Sturmian separable expansion approach. This solution method
provides an exact description of the threshold phenomena. We have found
several new $S$-wave resonances in the $e^-e^+e^-$ system 
in the vicinity of thresholds.
\end{abstract}

\end{frontmatter}

\section{Introduction}

Certainly, to calculate resonances in three body atomic systems the 
methods based on complex coordinates are the most popular. 
The complex rotation of the coordinates turns the resonant behavior of the 
wave function into a bound-state-like asymptotic
behavior. Then, standard bound-state methods become applicable also for 
calculating resonances. The complex rotation of the coordinates 
does not change the discrete spectrum, 
the branch cut, corresponding to scattering states, however,
is rotated down onto the complex energy plane, and as a consequence, 
resonant states from the unphysical sheet become accessible. 

In practice bound state variational approaches are used, that results in
a discretization of the rotated continuum. By changing the rotation angle
the points corresponding to the continuum move, while those corresponding 
to discrete states, like bound and resonant states, stay.
This way one can determine resonance parameters.

However, in practical calculations
the points of the discretized continuum scatter around the rotated-down 
straight line. So, especially around thresholds it is not easy 
to decide whether a point is a resonance point or it belongs 
to the rotated continuum. 

Recently, we have developed a method  for calculating resonances in 
three-body Coulombic systems by solving homogeneous 
Faddeev-Merkuriev integral equations \cite{fm-book}
using the Coulomb-Sturmian separable expansion approach \cite{pdh1}.
As a test case, we calculated the resonances of the negative positronium
ion. We found all the $12$ resonances presented in Ref.\ \cite{ho} 
and observed good agreements in all cases.

With our method we succeeded in locating $10$ more
resonances in the same energy region, all of them are very close to the 
thresholds.
This is certainly due to the fact that in Ref.\ \cite{pdh1} 
the threshold behaviors are exactly taken into account. 
Unexpectedly we also observed that in the case of attractive 
Coulomb interactions the Faddeev-Merkuriev integral equations 
produce numerous resonant solutions of dubious origin. 
We tend to regard them as spurious resonances which come about 
from the somewhat arbitrary splitting of 
the potential in the three-body configuration space into short-range
and long-range terms, that is an inherent attribute of the theory.

Since the possible appearance of spurious states in the formalism is new and
surprising, for the sake of the better understanding of its mechanism,
in Section 2 we outline briefly 
the Faddeev-Merkuriev integral equation formalism.
In our particular case of the $e^-e^+e^-$ system two particles are identical 
and the set of Faddeev-Merkuriev integral equations can be reduced to an 
one-component equation.
In Section 3 we discuss the  spurious solutions.
In Section 4 the Coulomb-Sturmian separable expansion method is applied
to the one-component integral equation.
In Section 5 we present our new calculations for the $L=0$  resonances
of the $e^-e^+e^-$ system  and compare them with the
results of the complex scaling calculations of Ref.\ \cite{ho}.

\section{Faddeev-Merkuriev integral equations}

The Hamiltonian of a three-body Coulombic system is given by
\begin{equation}
H=H^0 + v_1^C+ v_2^C + v_3^C,
\label{H}
\end{equation}
where $H^0$ is the three-body kinetic energy
operator and $v_\alpha^C$ denotes the
Coulomb potential in the subsystem $\alpha$, with $\alpha=1,2,3$.
We use throughout the usual configuration-space Jacobi coordinates
$x_\alpha$  and $y_\alpha$. Thus  $v_\alpha^C$, the potential between
particles $\beta$ and $\gamma$, depends on $x_\alpha$.

The  Hamiltonian (\ref{H}) is defined in the three-body 
Hilbert space. The three-body kinetic energy, when the center-of-mass
motion is separated, is given by
\begin{equation}
H^0=h^0_{x_\alpha}+h^0_{y_\alpha}=h^0_{x_\beta}+h^0_{y_\beta}=
h^0_{x_\gamma}+h^0_{y_\gamma},
\end{equation}
where $h^0$ is the two-body kinetic energy.
The two-body potential operators are formally
embedded in the three-body Hilbert space
$v^C = v^C (x) {\bf 1}_{y}$,
where ${\bf 1}_{y}$ is a unit operator in the two-body Hilbert space
associated with the $y$ coordinate.

\begin{figure}
\includegraphics[width=0.75\textwidth,angle=-90]{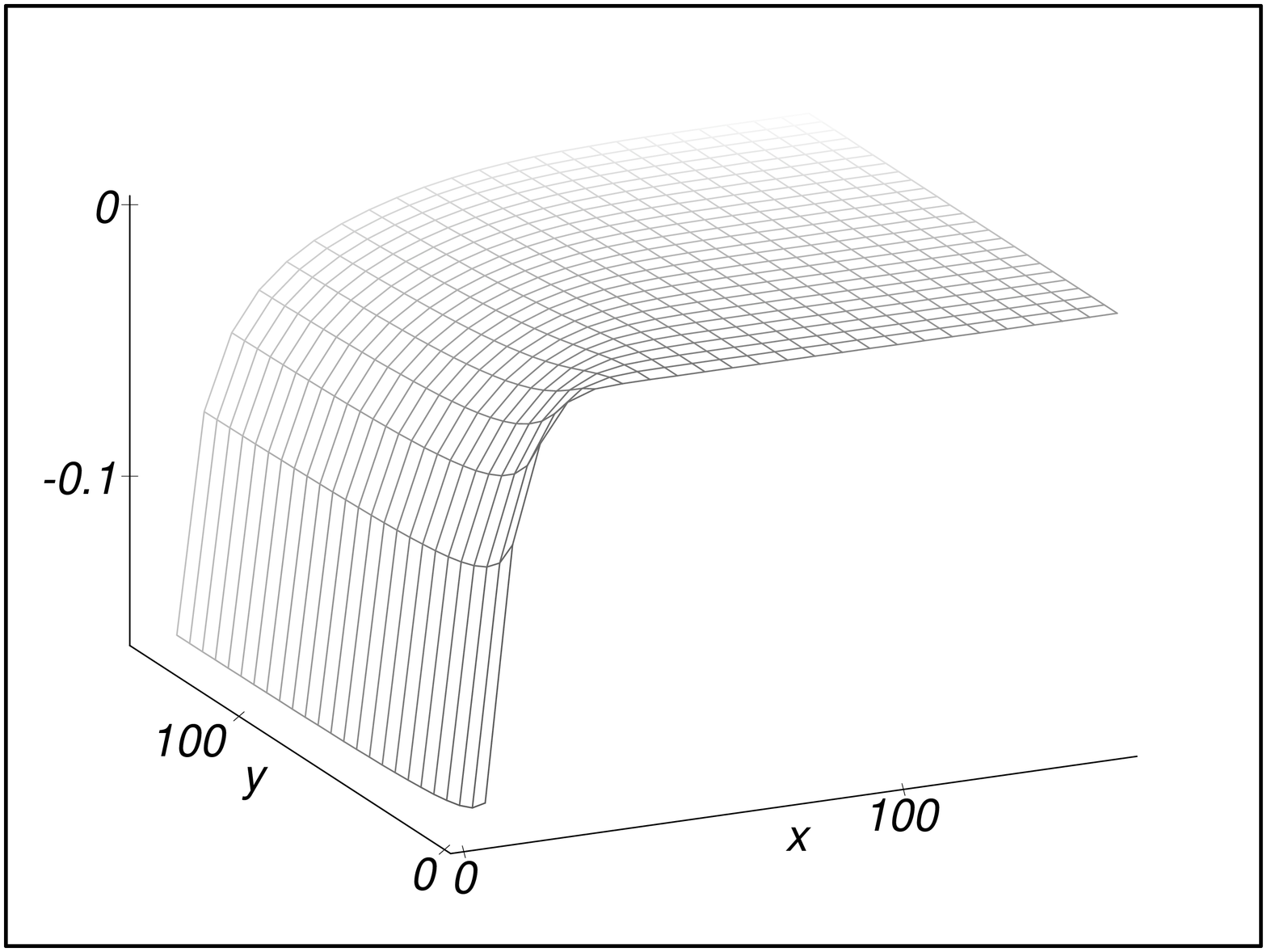}
\caption{Typical shape of $v^{(s)}$ for attractive Coulomb
potentials. }
\label{vs}
\end{figure}

\begin{figure}
\includegraphics[width=0.75\textwidth,angle=-90]{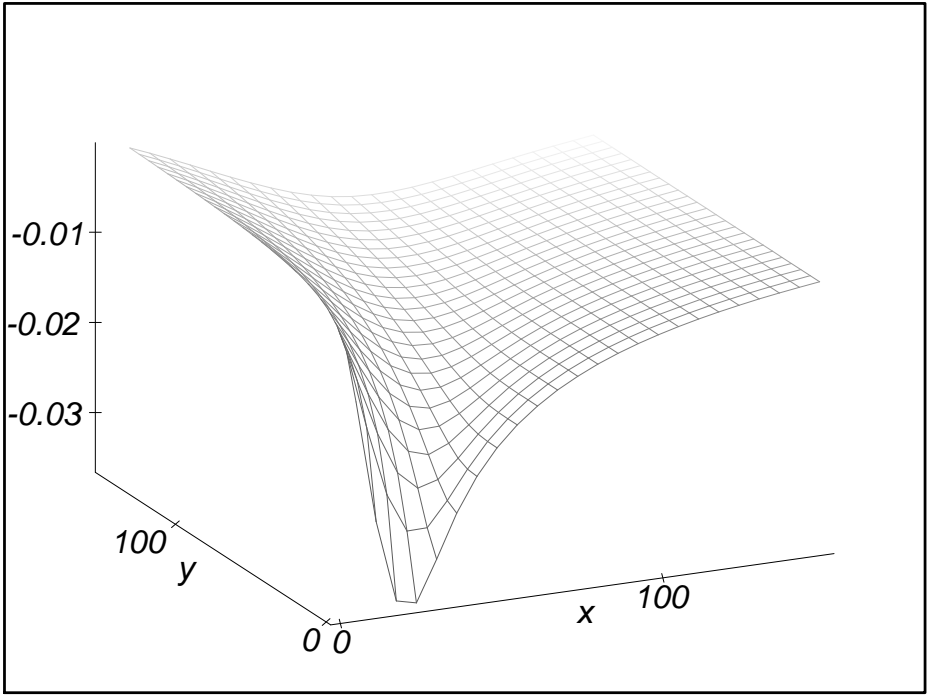}
\caption{Typical shape of $v^{(l)}$ for attractive Coulomb
potentials.}
\label{vl}
\end{figure}

Merkuriev proposed \cite{fm-book} to split the Coulomb interaction  in 
the three-body configuration space into
short- and long-range terms 
\begin{equation}
v_\alpha^C =v_\alpha^{(s)} +v_\alpha^{(l)} ,
\label{pot}
\end{equation}
where the short- and long-range parts are defined via a splitting
function:
\begin{eqnarray}
v_\alpha^{(s)} & = & v_\alpha^C \zeta (x_\alpha,y_\alpha) \\
v_\alpha^{(l)} & = & v_\alpha^C \left[ 1- \zeta (x_\alpha,y_\alpha)\right].
\label{potsl}
\end{eqnarray}
The splitting function $\zeta$
is defined in such a way that
\begin{equation}
\lim_{x,y \to \infty} \zeta (x,y) =
\left\{ 
\begin{array}{ll}
1, &  \mbox{if}\  |x| <  x^0 ( 1 +|y|/ y^0)^{1/\nu}, \\
0,  & \mbox{otherwise,}
\end{array}
\right.
\end{equation}
where $x^0, y^0 >0$ and $\nu > 2$.
Usually the functional form
\begin{equation}
\zeta (x,y) =  2/\left\{1+ 
\exp \left[ {(x/x^0)^\nu}/{(1+y/y^0)} \right] \right\},
\label{oma1}
\end{equation}
is used. 
So, the separation into short- and long-range parts is made along a 
parabola-like curve over the $\{x,y\}$ plane.
Typical shapes for $v^{(s)}$ and $v^{(l)}$ 
are shown in Figures \ref{vs} and \ref{vl}, respectively.

In atomic three-particle systems the sign of the charge of
two particles are always identical. Let us denote in our $e^-e^+e^-$ system
the two electrons by $1$ and $2$, and the positron by $3$.
In this case  $v_3^C$, the interaction between the two electrons,
is a repulsive Coulomb
potential which does not support two-body bound states. Therefore the entire
$v_3^C$ can be considered as long-range potential.
With splitting (\ref{pot}) the Hamiltonian can formally
be written in a form which looks like an usual
three-body Hamiltonian with two short-range potentials
\begin{equation}
H = H^{(l)} + v_1^{(s)}+  v_2^{(s)},
\label{hll}
\end{equation}
where the long-range Hamiltonian is defined as
\begin{equation}
H^{(l)} = H^0 + v_1^{(l)}+ v_2^{(l)}+ v_3^C.
\end{equation}
Then, the Faddeev method is applicable and, in this particular case, 
results in a splitting of the wave function into two components
\begin{equation}
|\Psi \rangle = |\psi_1 \rangle +
|\psi_2 \rangle,
\end{equation}
with components defined by 
\begin{equation}
|\psi_\alpha \rangle = G^{(l)} (z) v_\alpha^{(s)} |\Psi \rangle,
\label{filter}
\end{equation}
where $\alpha=1,2$ and $G^{(l)} (z) =(z-H^{(l)})^{-1}$, $z$ is a complex number.

In the case of bound and resonant states the wave-function components satisfy 
the homogeneous two-component Faddeev-Merkuriev integral equations
\begin{eqnarray}
 |\psi_1 \rangle &=& G_1^{(l)} (z)
v^{(s)}_1 |\psi_2 \rangle 
\label{fn-eq1} \\
 |\psi_2 \rangle &=& G_2^{(l)} (z)
v^{(s)}_2  |\psi_1 \rangle 
\label{fn-eq2} 
\end{eqnarray}
at real and complex energies, respectively.
Here $G^{(l)}_\alpha$ is the resolvent of the channel 
long-ranged Hamiltonian
$H^{(l)}_\alpha = H^{(l)} + v_\alpha^{(s)}$,
$G^{(l)}_\alpha(z)=(z-H^{(l)}_\alpha)^{-1}$.
Merkuriev has proved that Eqs.\ (\ref{fn-eq1}-\ref{fn-eq2}) 
possess compact kernels, and this property remains valid also for
complex energies $z=E-i\Gamma/2$, $\Gamma > 0$.

Further simplification can be achieved if we take into account that the
two electrons, particles $1$ and $2$, are identical and indistinguishable. 
Then, the Faddeev components 
$| \psi_1 \rangle$ and $| \psi_2 \rangle$, in their own natural Jacobi
coordinates, have the same functional forms
\begin{equation}
\langle x_1 y_1 | \psi_1 \rangle = \langle x_2 y_2 | \psi_2 \rangle
= \langle x y | \psi \rangle.
\end{equation}
On the other hand
\begin{equation}
| \psi_2 \rangle = p {\mathcal P}| \psi_1 \rangle,
\end{equation}
where ${\mathcal P}$ is the operator for the permutation of indexes
$1$ and $2$ and $p=\pm 1$ are eigenvalues of ${\mathcal P}$.
Therefore we can  determine $| \psi \rangle$ from the first equation
only 
\begin{equation} \label{fmp}
| \psi \rangle =  G_1^{(l)} v_1^{(s)} p {\mathcal P} | \psi \rangle.
\end{equation}
It should be emphasized, that so far we did not make any approximation,
and although this integral equation 
has only one component, yet it gives full account both
of asymptotic and symmetry properties of the system.

\section{Spurious resonance solutions}

Let us suppose for a moment that the $y^0$ parameter in $\zeta(x_1,y_1)$ 
is infinite.
Then $\zeta$ would not depend on $y_1$, 
the separation of the potential into 
short and long range parts would go along a straight line, and 
$v_1^{(l)}(x_1,y_1)$
would be like a valley in the $y_1$ direction. 
The potential 
$v_1^{(l)}(x_1,y_1) $, which is  Coulomb-like along $x_1$, would support
infinitely many bound states and, 
as free motion along the coordinate $y_1$ would be possible,
the Hamiltonian $H^0+v_1^{(l)}$
would have infinitely many two-body channels.

If, however, $y^0$ is finite, the straight line along the $y_1$ direction
becomes a parabola-like curve and the valley, as $y_1$ goes to infinity,
gets broader and broader and shallower and shallower, and finally 
disappears (see Fig \ref{vl}). 
As the valley gets closed the continuum of $H^0+v_1^{(l)}$ 
associated with the $y_1$ coordinate becomes discretized. 
So, if $y^0$ is finite,
$H^0+v_1^{(l)}$ have infinitely many bound states.
Similar analysis is valid also for $H^0+v_2^{(l)}$, and consequently, also
$H^{(l)}$ has infinitely many bound states. 

This, however,
due to (\ref{filter}), can lead to spurious solutions. If, at some energy,
$H^{(l)}$ has bound state $G^{(l)}$ has pole. Then, $|\psi_\alpha \rangle$
in Eq.\  (\ref{filter}), irrespective of $v^{(s)}_\alpha$, 
can be finite even if $|\Psi \rangle$ is infinitesimal.
These solutions are called spurious solutions: although the Faddeev 
components are not identically zero, but their sum vanishes.

Let us examine now the spectral properties of the Hamiltonian
\begin{equation} 
H_1^{(l)}=H^{(l)}+v^{(s)}_1=H^0+v_1^C+v_2^{(l)}+v_3^C.
\end{equation}
The three-body potential $v_2^{(l)}$ is attractive and constructed so that 
$v_2^{(l)}(x_2,y_2)$ vanishes as $y_2$ tends to infinity. 
Therefore, there are no
two-body channels associated with fragmentation $2$, and of course neither 
with fragmentation $3$, the Hamiltonian 
$H_1^{(l)}$ has only $1$-type two-body asymptotic channels.

What happens to the bound states associated with $v_2^{(l)}$? They are
embedded in the the continuum of $H^0+v_1^C$, and become resonant states.
So, by solving the Faddeev-Merkuriev integral 
equations, at some complex energies, we can encounter spurious solutions. 
These spurious solutions are 
not related to the Hamiltonian $H$, but rather only to some auxiliary 
potentials coming from the splitting procedure. 
Consequently, the spurious resonances should be sensitive to the parameters of 
$\zeta$, while the physical resonances not. 
This way we can easily distinguish between physical and spurious 
resonance solutions.

\section{Coulomb-Sturmian potential separable expansion approach}

We solve Eq.\ (\ref{fmp})
by using the Coulomb--Sturmian separable expansion approach \cite{pzwp}.
The Coulomb-Sturmian (CS) functions are defined by
\begin{equation}
\langle r|n l \rangle =\left[ \frac {n!} {(n+2l+1)!} \right]^{1/2}
(2br)^{l+1} \exp(-b r) L_n^{2l+1}(2b r),  \label{basisr}
\end{equation}
with $n$ and $l$ being the radial and
orbital angular momentum quantum numbers, respectively, and $b$ is the size
parameter of the basis.
The CS functions $\{ |n l \rangle \}$
form a biorthonormal
discrete basis in the radial two-body Hilbert space; the biorthogonal
partner is defined  by $\langle r |\widetilde{n l}\rangle=
\langle r |{n l}\rangle/r$. 
Since the three-body Hilbert space is a direct product of two-body
Hilbert spaces an appropriate basis
can be defined as the
angular momentum coupled direct product of the two-body bases 
\begin{equation}
| n \nu  l \lambda \rangle_1 =
 | n  l \rangle_1 \otimes |
\nu \lambda \rangle_1, \ \ \ \ (n,\nu=0,1,2,\ldots),
\label{cs3}
\end{equation}
where $| n  l \rangle_1$ and $|\nu \lambda \rangle_1$ are associated
with the coordinates $x_1$ and $y_1$, respectively.
With this basis the completeness relation
takes the form (with angular momentum summation
implicitly included)
\begin{equation}
{\bf 1} =\lim\limits_{N\to\infty} \sum_{n,\nu=0}^N |
 \widetilde{n \nu l \lambda} \rangle_1 \;\mbox{}_1\langle
{n \nu l \lambda} | =
\lim\limits_{N\to\infty} {\bf 1}^{N}_1.
\end{equation}
Note that similar bases can be constructed for fragmentations $2$ and
$3$ as well.

We make the following approximation on Eq.\ (\ref{fmp}) 
\begin{equation} \label{fmpa}
| \psi \rangle =  G_1^{(l)}
{\bf 1}^{N}_1 v_1^{(s)} p {\mathcal P} {\bf 1}^{N}_1
| \psi \rangle,
\end{equation}
i.e.\ the operator 
$v_1^{(s)} p {\mathcal P}$ in the three-body
Hilbert space is approximated by a separable form, viz.
\begin{eqnarray}
v_1^{(s)}p {\mathcal P} &  = &
\lim_{N\to\infty} {\bf 1}^{N}_1 v_1^{(s)} p {\mathcal P}  {\bf 1}^{N}_1
\nonumber \\ 
& \approx &  {\bf 1}^{N}_1 v_1^{(s)} p {\mathcal P} {\bf 1}^{N}_1 
 \approx   \sum_{n,\nu ,n^{\prime },
\nu ^{\prime }=0}^N|\widetilde{n\nu l \lambda}\rangle_1 \;
\underline{v}_1^{(s)}
\;\mbox{}_1 \langle \widetilde{n^{\prime}
\nu ^{\prime} l^{\prime} \lambda^{\prime}}|,  \label{sepfep}
\end{eqnarray}
where $\underline{v}_1^{(s)}=\mbox{}_1 \langle n\nu l \lambda|
v_1^{(s)} p {\mathcal P}  
|n^{\prime }\nu ^{\prime} l^{\prime} \lambda^{\prime}\rangle_1$.
The compactness of the equation and the completeness of the basis guarantee
the convergence of the method.
Utilizing the properties of the exchange operator ${\mathcal P}$
these matrix elements can be written in the form $\underline{v}_1^{(s)}= 
p\times (-)^{l^{\prime}} \; \mbox{}_1 \langle n\nu l \lambda| 
v_1^{(s)}|n^{\prime }\nu ^{\prime} l^{\prime} \lambda^{\prime}\rangle_2$.

With this approximation, the solution of Eq.\ (\ref{fmp})
turns into solution of matrix equations for the component vector
$\underline{\psi} =
 \mbox{}_1 \langle \widetilde{ n\nu l \lambda} | \psi  \rangle$
\begin{equation}
 \{ [ \underline{G}^{(l)}_1(z)]^{-1} - \underline{v}^{(s)}_1 \} 
\underline{\psi} =0,
\end{equation}
where  $\underline{G}_1^{(l)}=\mbox{}_1 \langle \widetilde{
n\nu l\lambda} |G_1^{(l)}|\widetilde{n'\nu' l' \lambda' }\rangle_1$. 
A unique solution exists if
and only if
\begin{equation}
D(z)\equiv 
\det \{ [ \underline{G}^{(l)}_1 (z)]^{-1} - \underline{v}^{(s)}_1 \} =0.
\label{fdet}
\end{equation}

The Green's operator  $\underline{G}_1^{(l)}$  
is related to the Hamiltonian $H_1^{(l)}$, 
which is still a three-body Coulomb
Hamiltonian, seems to be as complicated as $H$ itself.
However, $H_1^{(l)}$ has only $1$-type asymptotic channels, with
asymptotic Hamiltonian
\begin{equation} \label{htilde}
\widetilde{H}_1 = H^{0}+v_1^C.
\end{equation}
Therefore, in the spirit of the three-potential formalism \cite{phhky},
$\underline{G}_1^{(l)}$ can be linked to the matrix elements of
$\widetilde{G}_1(z)=(z-\widetilde{H}_1)^{-1}$ 
via solution of a Lippmann-Schwinger equation,
\begin{equation}
(\underline{G}^{(l)}_1)^{-1}= 
(\underline{\widetilde{G}}_1)^{-1} -
\underline{U}_1,
\label{gleq}
\end{equation}
where 
${\underline{\widetilde{G}}_1}  =
 \mbox{}_1\langle \widetilde{n \nu l \lambda} | 
 \widetilde{G}_1 |
 \widetilde{ n^{\prime}\nu^{\prime}l^{\prime}{\lambda}^{\prime}} \rangle_1 $
and ${\underline{U}_1} =
 \mbox{}_1\langle n\nu l \lambda | (v_2^{(l)}+v_3^C) | n^{\prime}\nu^{\prime}
l^{\prime}{\lambda}^{\prime}\rangle_1$.

The most crucial point in this procedure is the
 calculation of the matrix elements
$\underline{\widetilde{G}}_1$, since the  potential
matrix elements $\underline{v}^{(s)}_{1}$ and
$\underline{U}_1$ can always be evaluated numerically by making use of
the transformation of Jacobi coordinates \cite{bb}.
The Green's operator $\widetilde{G}_1$
is a resolvent of the sum of two commuting Hamiltonians,
$\widetilde{H}_1 = h_{x_1}+h_{y_1}$,
where $h_{x_1}=h^0_{x_1}+v_1^C(x_1)$ and
$h_{y_1}=h^0_{y_1}$,
which act in different two-body Hilbert spaces.
Thus, $\widetilde{G}_1$ can be given by 
a convolution integral of two-body Green's operators, i.e.
\begin{equation}
\widetilde{G}_1 (z)=
 \frac 1{2\pi {i}}\oint_C
dz' \,g_{x_1}(z-z')\;g_{y_1}(z'),
 \label{contourint}
\end{equation}
where
$g_{x_1}(z)=(z-h_{x_1})^{-1}$  and
$g_{y_1}(z)=(z-h_{y_1})^{-1}$.
The contour $C$ should be taken  counterclockwise
around the continuous spectrum of $h_{y_1}$
such a way that $g_{x_1}$ is analytic on the domain encircled
by $C$.

\begin{figure}
\includegraphics[width=0.80\textwidth]{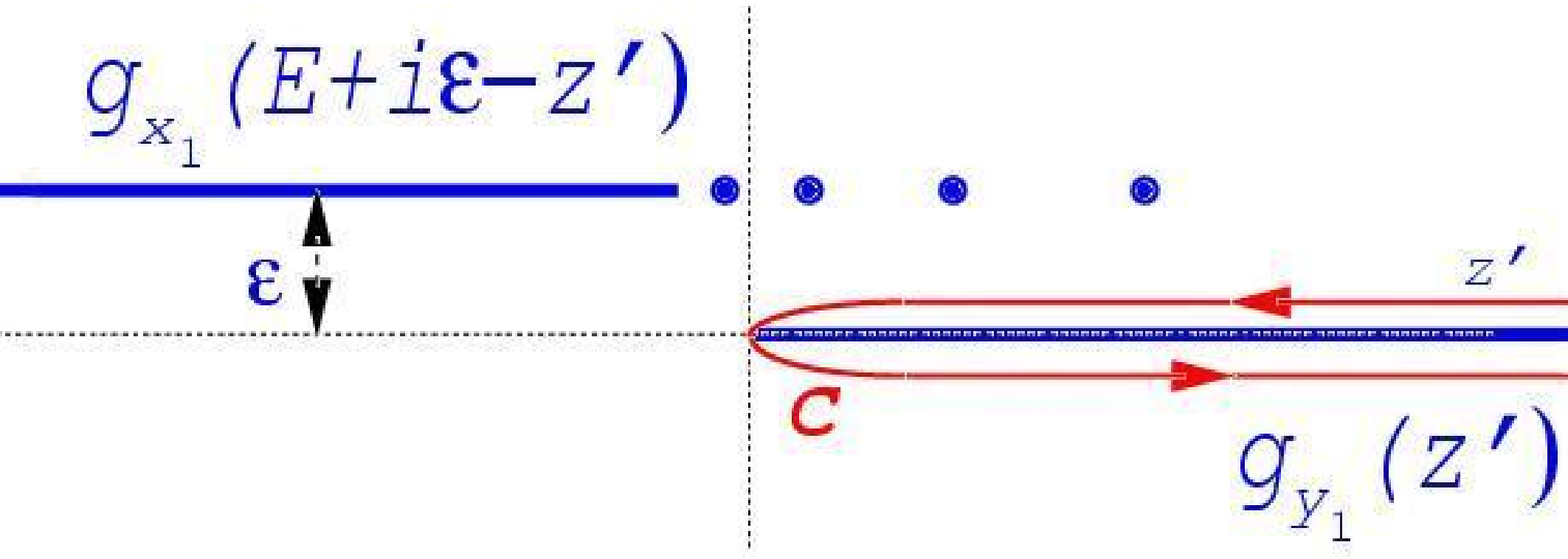}
\caption{Analytic structure of $g_{x_1}(E+{\mathrm{i}}\varepsilon-z')\;
g_{y_1}(z')$ as a function of $z'$, $\varepsilon>0$. The Green's operator
$g_{y_1}(z')$ has a branch-cut on the $[0,\infty)$ interval, while
$g_{x_1}(E+{\mathrm{i}}\varepsilon-z')$ has a branch-cut on the 
$(-\infty,E+{\mathrm{i}}\varepsilon]$ interval and infinitely many 
poles accumulated at $E+{\mathrm{i}}\varepsilon$ (denoted by dots).
The contour $C$ encircles the branch-cut of
$g_{y_1}$. In the $\varepsilon \to 0$ limit the singularities
of $g_{x_1}(E+{\mathrm{i}}\varepsilon -z')$ would penetrate
into the area covered by $C$. }
\label{fig1}
\end{figure}

\begin{figure}
\includegraphics[width=0.80\textwidth]{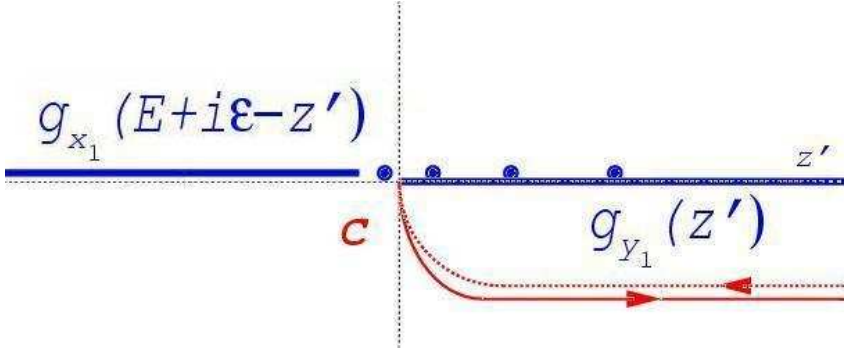}
\caption{The contour of Fig.\ \ref{fig1} is
deformed analytically such that a part of it goes on the unphysical
Riemann-sheet of $g_{y_1}$ (drawn by broken line) and the other part detoured
away from the cut. Now, the contour avoids the singularities of 
$g_{x_1}(E+{\mathrm{i}}\varepsilon-z')$ even in the
 $\varepsilon \to 0$ limit.}
\label{fig2}
\end{figure}

\begin{figure}
\includegraphics[width=0.80\textwidth]{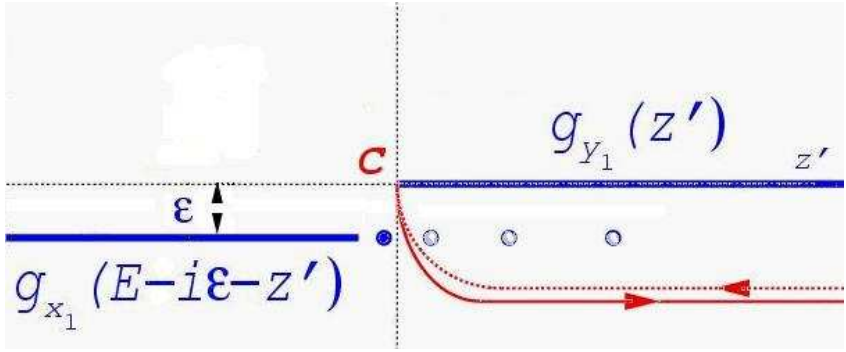}
\caption{In the $\varepsilon < 0$ case, what is needed to calculate resonances,
the poles of $g_{x_1}$ which lie above the branch point of $g_{y_1}$ go
to the unphysical sheet of  $g_{y_1}$ (denoted by shaded points), 
while the others remain on the physical one. }
\label{fig3}
\end{figure}

In time-independent scattering theory the Green's operator has a branch-cut
singularity at scattering energies. In our formalism $\widetilde{G}_1 (E)$
should be understood as 
$\widetilde{G}_1 (E)=\lim_{\varepsilon\to 0} \widetilde{G}_1 
(E +{\mathrm{i}}\varepsilon)$, with $\varepsilon > 0$, and $E < 0$, since
in this work we are considering resonances below the three-body breakup 
threshold. To calculate resonant states $\widetilde{G}_1 
(E +{\mathrm{i}}\varepsilon)$ has to be continued
analytically to negative $\varepsilon$ values.
Before doing that  let us examine 
$\widetilde{G}_1 (E +{\mathrm{i}}\varepsilon)$ with $\varepsilon > 0$.
Now, the spectra of 
$g_{y_1}(z')$ and $g_{x_1}(E +{\mathrm{i}}\varepsilon-z')$
are well separated and
the spectrum of $g_{y_1}$ can easily be encircled such that the spectrum
of $g_{x_1}$ does not penetrate into the encircled area (Fig.\ \ref{fig1}).
Next, the contour $C$ is deformed analytically
in such a way that the upper part descends into the unphysical
Riemann sheet of $g_{y_1}$, while
the lower part of $C$ is detoured away from the cut
(Fig.\ \ref{fig2}). The contour still
encircles the branch cut singularity of $g_{y_1}$,
but in the  $\varepsilon\to 0$ limit it now
avoids the singularities of $g_{x_1}$.
Moreover, by continuing to negative values of  $\varepsilon$, in order that
we can calculate resonances, the  pole singularities of
$g_{x_1}$ move onto the second Riemann sheet of $g_{y_1}$  (Fig.\ \ref{fig3}). 
Thus, the mathematical conditions for
the contour integral representation of $\widetilde{G}_1 (z)$ in
Eq.~(\ref{contourint}) can be fulfilled also for complex energies
with negative imaginary parts.
In this respect there is only a gradual difference between the
bound- and resonant-state calculations. Now,
the matrix elements $\underline{\widetilde{G}}_\alpha$
can be cast in the form
\begin{equation}
\widetilde{\underline{G}}_1 (z)=
 \frac 1{2\pi {i}}\oint_C dz' \,\underline{g}_{x_1}(z-z')\;
\underline{g}_{y_1}(z'),
\label{contourint2}
\end{equation}
where the corresponding CS matrix elements of the two-body Green's operators in
the integrand are known analytically for all complex energies (see \cite{phhky}
and references therein).
It is also evident that all the thresholds, corresponding to the 
poles of $\underline{g}_{x_1}$, are at the right location and therefore 
this method is especially suited to study near-threshold resonances.

\section{ $S$-wave resonances in positronium ion}

To calculate resonances we have to find the complex zeros
of the Fredholm determinant (\ref{fdet}). To locate them we calculate
$D(z)$ along the real energy axis with small step size.
In the vicinity of zeros the $D(z)$ exhibits violent
changes. Then we have a good starting point for the zero search.
This way we can easily find resonances, at least the narrow ones.
To find broad resonances one should proceed by calculating
$D(E-i\varepsilon)$ with finite $\varepsilon >0$.

For the parameters of the splitting function
we take $y^0=50\mbox{a}_0$ and $\nu=2.1$, $a_0$ is the Bohr radius. To
select out the physical solutions we vary the short- and long-range
potentials by taking $x^0=18\mbox{a}_0,25\mbox{a}_0$ and  
$30\mbox{a}_0$, respectively.
We can also vary the potentials by
adding and subtracting a short-range term
\begin{eqnarray}
v_1^{(s)} & = & v_1^C \zeta (x_1,y_1) + v_0  \zeta (x_1,y_1) \\
v_1^{(l)} & = & v_1^C \left[ 1- \zeta (x_1,y_1)\right]  -
 v_0  \zeta (x_1,y_1),
\label{potslu}
\end{eqnarray}
respectively, while keeping a fixed $x^0=18\mbox{a}_0$ value. 
This new kind of splitting goes beyond the
Merkuriev's original suggestion, but since the character
of $v^{(s)}$ and $v^{(l)}$ remained the same all the nice
properties of the original Faddeev-Merkuriev equations are retained.
In these calculations we used $v_0=0, 0.01, 0.02$ values.

We found that some solutions, especially their widths, are very
sensitive to the change of either $x^0$ or $v_0$ parameters.
We regard them spurious solutions. Those solutions, given in
Table I, were stable against all changes of parameters and we
can consider them as physical resonances.
We recovered all the resonances presented in Ref.\ \cite{ho} with very
good agreements, but besides that we found $10$ more resonances,
all of them are in the vicinity of some thresholds.

\begin{table}
\begin{center}
\begin{tabular}{l|llll}
 State & \multicolumn{2}{c}{Ref.\ \cite{ho}} & \multicolumn{2}{c}{Our
work}
 \\ \hline
$^1S^e$ & \multicolumn{1}{c}{$-E_r$}  & \multicolumn{1}{c}{$\Gamma$}
 & \multicolumn{1}{c}{$-E_r$}  & \multicolumn{1}{c}{$\Gamma$}  \\
\hline
  &  0.1520608 & 0.000086 &  0.15192   & 0.0000426   \\
  &  0.12730   & 0.00002  &  0.12727   & 0.0000084   \\
  &            &          &  0.1251    & 0.000002    \\
  &  0.070683  & 0.00015  &  0.70666   & 0.000074   \\
  &  0.05969   & 0.00011  &  0.059682  & 0.0000526  \\
  &            &          &  0.0564     & 0.00003     \\
  &  0.04045   & 0.00024  &  0.40426   & 0.0001294  \\
  &  0.0350    & 0.0003   &  0.0350206 & 0.00013   \\
  &  0.03463   & 0.00034  &  0.0346234 & 0.0001586   \\
  &            &          &  0.03263   & 0.0001    \\
  &            &          &  0.03158    & 0.00007    \\
  &  0.0258    & 0.00045  &  0.02606   & 0.000104   \\
  &  0.02343   & 0.00014  &  0.023428  & 0.0000436  \\ \hline
$^3S^e$ & \multicolumn{1}{c}{$-E_r$}  & \multicolumn{1}{c}{$\Gamma$}
 & \multicolumn{1}{c}{$-E_r$}  & \multicolumn{1}{c}{$\Gamma$}  \\ \hline
  &  0.12706   & 0.00001   &  0.127050  & 0.0000000028   \\
  &            &           &  0.1251    & 0.000000001  \\
  & 0.05873    & 0.00002   &  0.05872   & 0.0000001852  \\
  &            &           &  0.0561    & 0.0000002     \\ 
  &            &           &  0.0553    & 0.0001          \\
  &  0.03415   & 0.00002   &  0.0342018 & 0.00000070       \\
  &            &           &  0.03237   & 0.0000011      \\
  &            &           &  0.03172   & 0.0000002      \\
  &            &           &  0.031035  & 0.0000938
\end{tabular}
\end{center}
\caption{
$L=0$ resonances of $\mbox{Ps}^-$. The energies and
widths are
expressed in Rydbergs.
\label{tab_b}}
\end{table}

\section{Conclusions}

In this article we pointed out that in the Faddeev-Merkuriev 
integral equation approach of the three-body Coulomb problem
the complex-energy spectrum is contaminated with
spurious solutions. The spurious solutions, however, are sensitive
to the splitting of the potential into short- and long-range terms.
This offers an easy way to select the physical solutions.

We solved the integral equations
by using the Coulomb-Sturmian separable expansion approach. 
This method gives an exact description of the threshold phenomena, thus
the method is ideal for studying close-to-threshold resonances.
In the $e^- e^+ e^-$ system we located $10$ new resonances, all of them
are in the close vicinity of some threshold.

This work has been supported by the NSF Grant No.Phy-0088936
and OTKA Grants under Contracts No.\ T026233 and No.\ T029003.

\end{document}